\documentclass[]{elsart}

\usepackage{graphicx}
\usepackage{dcolumn}
\usepackage{amsmath}
\usepackage[]{fontenc}
\usepackage{longtable}

\begin{document}

\begin{frontmatter}
\title{Measurements of Relative Branching Ratios of $\Lambda _{c}^{+}$ Decays
into States Containing $\Sigma$ }

The FOCUS Collaboration

\author[ucd]{J.~M.~Link}
\author[ucd]{M.~Reyes}
\author[ucd]{P.~M.~Yager}
\author[cbpf]{J.~C.~Anjos}
\author[cbpf]{I.~Bediaga}
\author[cbpf]{C.~G\"obel}
\author[cbpf]{J.~Magnin}
\author[cbpf]{A.~Massafferri}
\author[cbpf]{J.~M.~de~Miranda}
\author[cbpf]{I.~M.~Pepe}
% ...

\address[ucd]{University of California, Davis, CA 95616}
\address[cbpf]{Centro Brasileiro de Pesquisas F\'isicas, Rio de Janeiro, RJ, Brasil}
\address[cinv]{CINVESTAV, 07000 M\'exico City, DF, Mexico}
\address[cu]{University of Colorado, Boulder, CO 80309}
\address[fnal]{Fermi National Accelerator Laboratory, Batavia, IL 60510}
\address[fras]{Laboratori Nazionali di Frascati dell'INFN, Frascati, Italy I-00044}
\address[ui]{University of Illinois, Urbana-Champaign, IL 61801}
\address[iu]{Indiana University, Bloomington, IN 47405}
\address[korea]{Korea University, Seoul, Korea 136-701}
\address[kp]{Kyungpook National University, Taegu, Korea 702-701}
\address[milan]{INFN and University of Milano, Milano, Italy}
\address[nc]{University of North Carolina, Asheville, NC 28804}
\address[pavia]{Dipartimento di Fisica Nucleare e Teorica and INFN, Pavia, Italy}
\address[pr]{University of Puerto Rico, Mayaguez, PR 00681}
\address[sc]{University of South Carolina, Columbia, SC 29208}
\address[ut]{University of Tennessee, Knoxville, TN 37996}
\address[vu]{Vanderbilt University, Nashville, TN 37235}
\address[wisc]{University of Wisconsin, Madison, WI 53706}

\date{\today}

\begin{abstract}
We have studied the Cabibbo suppressed decay \( \Lambda _{c}^{+}\rightarrow \Sigma ^{+}K^{*0}(892) \)
and the Cabibbo favored decays \( \Lambda _{c}^{+}\rightarrow \Sigma ^{+}K^{+}K^{-} \),
\( \Lambda _{c}^{+}\rightarrow \Sigma ^{+}\phi  \) and \( \Lambda _{c}^{+}\rightarrow \Xi ^{*0}(\Sigma^{+}K^{-})K^{+} \) and
measured their branching ratios relative to \( \Lambda _{c}^{+}\rightarrow \Sigma ^{+}\pi ^{+}\pi ^{-} \)
to be \( (7.8\pm 1.8\pm 1.3)\% \), \( (7.1\pm 1.1\pm 1.1)\% \), \( (8.7\pm 1.6\pm 0.6)\% \) and 
\( (2.2\pm 0.6\pm 0.6)\% \), respectively. The first error is statistical
and the second is systematic. We also report two $90\%$ confidence level limits \( \Gamma (\Lambda
_{c}^{+}\rightarrow \Sigma ^{-}K^{+}\pi ^{+})/\Gamma (\Lambda _{c}^{+}\rightarrow \Sigma^{+}K^{*0}(892))<35\% \)
and \( \Gamma (\Lambda _{c}^{+}\rightarrow \Sigma ^{+}K^{+}K^{-})_{NR}/\Gamma(\Lambda_{c}^{+}\rightarrow \Sigma ^{+}\pi ^{+}\pi ^{-})<2.8\% \).
\end{abstract}

\end{frontmatter}

\section{{\normalsize Introduction}\normalsize }

Past experiments have reported results on non-leptonic branching fractions
of the lowest lying charmed baryon \( \Lambda _{c}^{+} \) \cite{Abe:2001mb,Avery:1993vj}.
In this paper we report on several \( \Lambda _{c}^{+} \) decay channels containing
a \( \Sigma  \) baryon in the final state. These measurements may be useful
in testing theoretical predictions of the contributions to inclusive decay amplitudes.
For instance, as pointed out by Guberina and Stefancic \cite{Guberina:2002fz}, direct
measurements of \( \Lambda _{c}^{+} \) singly Cabibbo suppressed decay rates 
can improve our theoretical understanding of the \( \Xi _{c}^{+} \) lifetime, which
can then be compared to recent high statistics measurements \cite{Link:2001qy}.

\section{{\normalsize Reconstruction}\normalsize }

This analysis uses data collected by the \mbox{FOCUS} experiment at
Fermilab during the 1996-97 fixed-target run and is based on a topological sample
of events with a charged Sigma hyperon plus two other charged particles emerging 
from the \( \Lambda _{c}^{+} \) decay vertex.

\mbox{FOCUS} is a photo-production experiment equipped with very precise vertexing
and particle identification detectors. The vertexing system is composed of
a silicon microstrip detector (TS) \cite{TS_desc} interleaved with segments of the BeO 
target and a second system of twelve microstrip planes (SSD) downstream
of the target. Beyond the SSD, five stations of multi-wire proportional chambers plus
two large aperture dipole magnets complete the charged particle tracking and momentum
measurement system. Three multi-cell, threshold
\v{C}erenkov counters discriminate between different particle hypotheses, namely
electrons, pions, kaons and protons. The \mbox{FOCUS} apparatus also contains
one hadronic and two electromagnetic calorimeters as well as two muon detectors. 

Events are selected using a candidate driven vertexing algorithm where the vector
components of the reconstructed decay particles define the charm flight direction. This
is used as a seed track to find the production vertex \cite{Driv_alg}. Using
this algorithm we determine the confidence level of the decay and production
vertices, and the significance of their separation. For each of the decay
modes analyzed, we require the primary vertex to have a confidence level greater
than \( 1\% \) and to contain at least two tracks other than the charm
seed track.

The \( \Sigma ^{-}  \! \rightarrow \!  n\pi ^{-}, \) \( \Sigma ^{+}\!
\rightarrow \! n\pi ^{+} \) and
\( \Sigma ^{+}\! \rightarrow \! p\pi ^{0} \) decays\footnote{%
Throughout this paper the charge conjugate state is implied unless explicitly
stated. Note that the \( \Sigma ^{-} \) is not the charge conjugate partner
of the \( \Sigma ^{+} \).
} are reconstructed using a kink algorithm \cite{Link:2001dj} where the properties
of the neutral particle in the decay are not detected, but rather inferred,
with a two-fold ambiguity in the momentum solution for some decays. Systematic
effects due to this ambiguity are reduced by normalizing to the decay mode 
\( \Lambda _{c}^{+}\! \rightarrow \! \Sigma ^{+}\pi ^{+}\pi ^{-} \), where the
same effect exists. To aid in fitting the mass distribution, for channels
containing a $\Sigma$, we implement a double Gaussian to determine the yield
of signal events. By double Gaussian we mean two Gaussian shapes with separate
amplitudes, means and widths. In this paper we will discuss the 
$\Sigma^+ \pi^+ \pi^-$, $\Sigma^+ K^+ \pi^-$, $\Sigma^- K^+ \pi^+$ and 
$\Sigma^+ K^+ K^-$ final states. In the $\Lambda_c^+ \rightarrow \Sigma^+ \pi^+ \pi^-$
mode we let all the parameters float while in the other lower statistics modes 
we fix some of the parameters to their Monte Carlo values.

\section{{\normalsize \protect\( \Lambda _{c}^{+}\! \rightarrow \! \Sigma ^{+}\pi ^{+}\pi ^{-}\protect \)
normalization mode}\normalsize }

The \( \Lambda _{c}^{+}\! \rightarrow \! \Sigma ^{+}\pi ^{+}\pi ^{-} \) mode is our
highest statistics decay containing a \( \Sigma ^{+} \) particle.
The events are selected requiring a detachment between primary and secondary
vertex divided by its error (\( l/\sigma _{l}\)) greater then \( 5.5 \). A
minimum \( \Lambda _{c}^{+} \) momentum cut of \( 50 \) GeV/$c$  is imposed,
as is a minimum secondary vertex confidence level of \( 10\% \). We also apply a cut
on the lifetime resolution, \( \sigma _{t}<120 \)\,fs for the run period where
we had a silicon detector (TS) in the target region (about $2/3$ of the events)
and \( \sigma _{t}<150 \)\,fs otherwise (NoTS). Further, we reject events which have
a lifetime greater than six times the \( \Lambda _{c}^{+} \) lifetime. 

\begin{figure}[htb!]
{\par\centering \resizebox*{8cm}{8cm}{\includegraphics{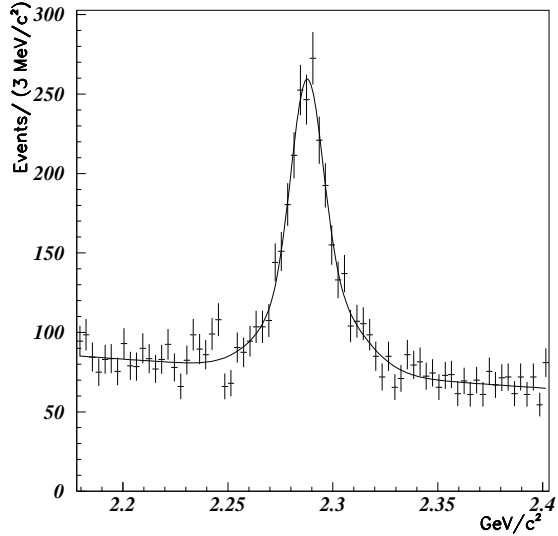}} \par}

\caption{\label{spp}\protect\( \Sigma ^{+}\pi ^{+}\pi ^{-}\protect \)
invariant mass distribution fit with a double Gaussian for the signal and a linear background.}
\end{figure}

We identify charged tracks using information from three \v{C}erenkov counters. The
algorithm \cite{Link:2001pg} makes use of the on/off status of the \v{C}erenkov cells to
construct a likelihood for the hypothesis that a given track of a certain
momentum  generates the observed output. Once the likelihood for each hypothesis
``\( \alpha  \)'' (proton, kaon, pion, electron) has been computed we construct
a \( \chi ^{2} \)-like variable \( W_{\alpha }=-2\ln (Likelihood_{\alpha }) \).
We identify particles by comparing these variables.

Our pion consistency (PICON) requirement comes from comparing the minimum \( W_{\alpha } \) hypothesis
to the pion hypothesis. For example, the two pions from \( \Lambda _{c}^{+}\! \rightarrow \! \Sigma ^{+}\pi ^{+}\pi ^{-} \)
decay are required to satisfy PICON=\( \min \{W_{\alpha =e,\pi ,K,p}\}-W_{\pi }>-6 \).
Further, pions must not be identified as muons by the muon detector. For $\Sigma^+ \rightarrow 
p \pi^0$ decays, the proton is required to
satisfy $W_\pi - W_p > -3$ while for $\Sigma^+ \rightarrow n \pi^+$, the pion
must satisfy $W_p - W_\pi > -3$ and PICON $> -6$ and have a momentum greater
than 5\,GeV/$c$.

The \( \Sigma ^{+}\pi ^{+}\pi ^{-} \) mass distribution (Fig. \ref{spp}) is
fit with a double Gaussian and a linear background. We obtain a yield of \( 1706\pm 88 \)
events. 

\section{{\normalsize \label{skstar_section}\protect\( \Lambda _{c}^{+}\! \rightarrow \! \Sigma ^{+}K^{*0}(892)\protect \)
and \protect\( \Lambda _{c}^{+}\! \rightarrow \! \Sigma ^{-}K^{+}\pi ^{+}\protect \)
decay modes} }

\begin{figure}[htb!]
{\par\centering \resizebox*{6cm}{8cm}{\includegraphics{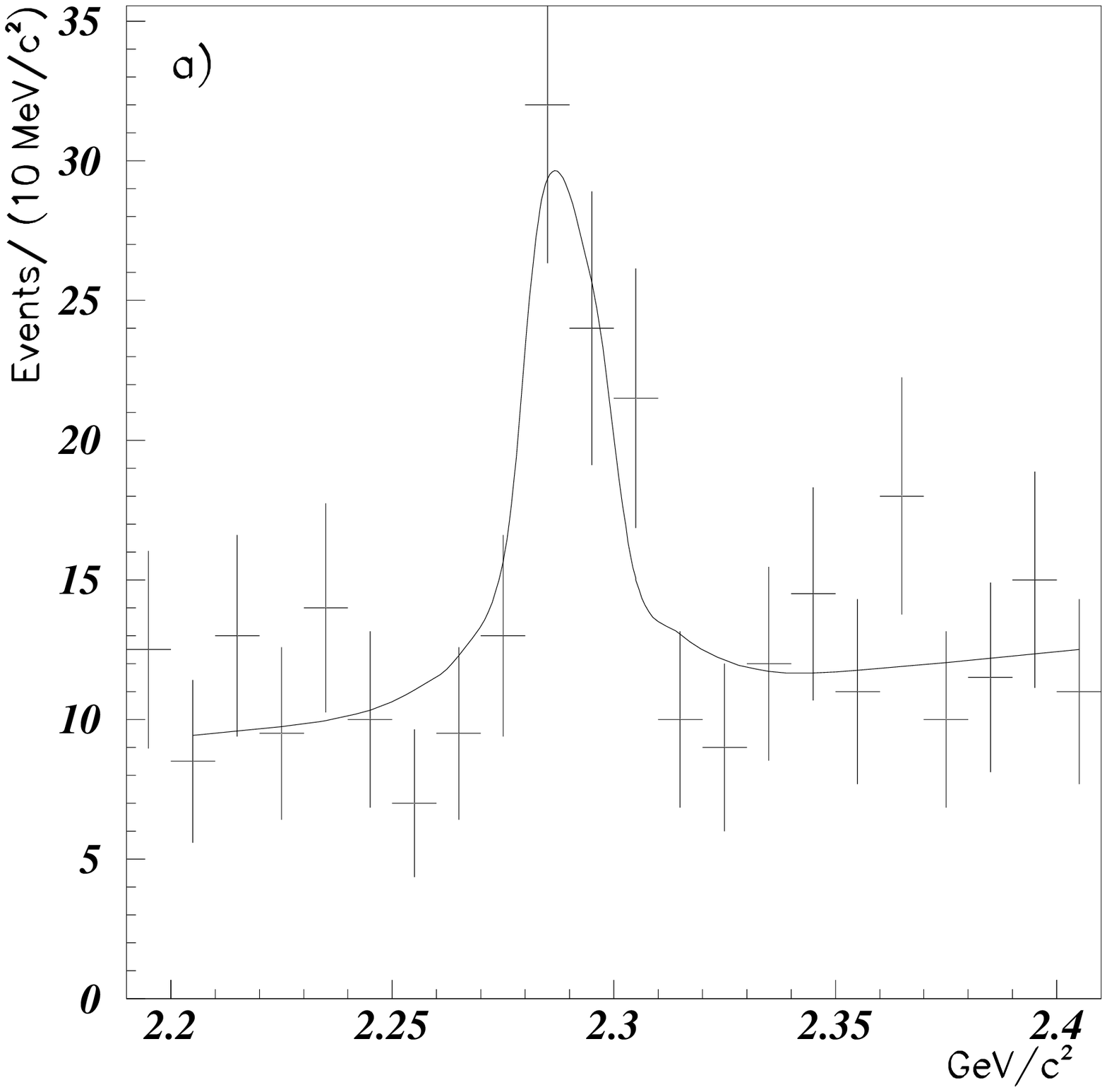}}  \resizebox*{6cm}{8cm}{\includegraphics{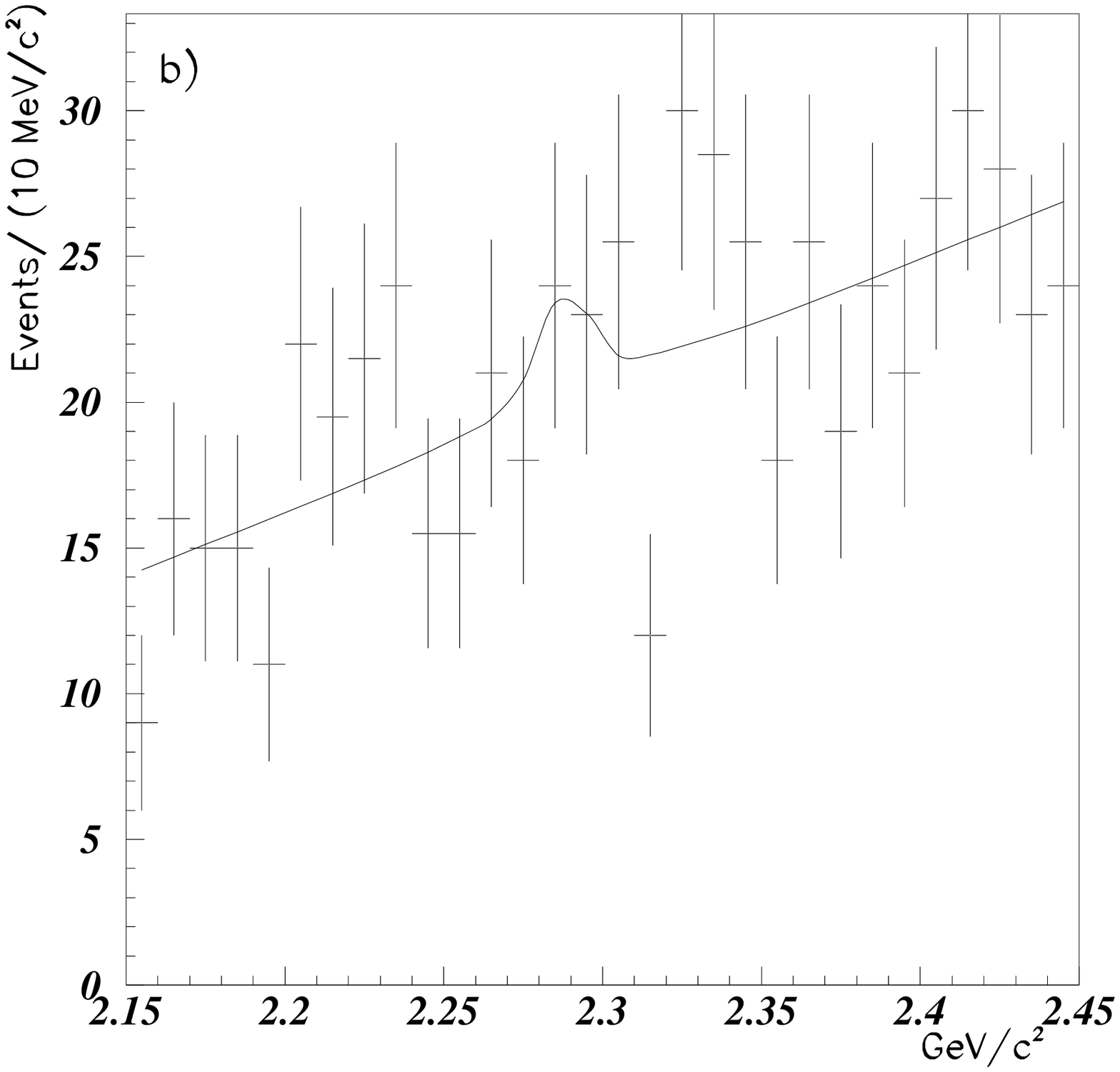}} \par}

\caption{\label{skstar}a) The \protect\( \Sigma ^{+}K^{*0}\protect \) invariant mass
distribution where the \protect\( K^{*0}(892)\protect \) is reconstructed in
the \protect\( K^{+}\pi ^{-}\protect \) channel. Only events in the $K^*(892)$
signal region are selected. b) Invariant mass distribution
for the final state \protect\( \Sigma ^{-}K^{+}\pi ^{+}\protect \). The distributions are fit
with a double Gaussian for the signal and a linear background.}
\end{figure}

\begin{figure}[htb!]
{\par\centering \resizebox*{6cm}{4cm}{\includegraphics{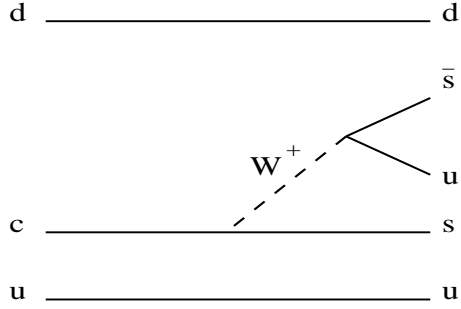}} \par}
\caption{\label{spectator}Spectator diagram for \protect\( \Lambda _{c}^{+}\! \rightarrow \! \Sigma ^{+}K^{*0}(892)\protect \). 
In this case no quark pair needs to be created.}
\end{figure}

The \( \Lambda _{c}^{+} \) is reconstructed in the decay channel \( \Sigma
^{+}K^{+}\pi ^{-} \). We fit the invariant mass distribution with and without
a mass cut (\( 832< \)\( M(K\pi ) \)\( <960 \) MeV/$c^{2}$) around the
$K^*(892)$ nominal value.
We find that most, if not all, of this channel occurs via \( \Sigma ^{+}K^{*0}(892) \). The absence 
of a non-resonant decay may also explain the absence of a signal in the \( \Lambda _{c}^{+} \)
decay to \( \Sigma ^{-}K^{+}\pi ^{+} \). In Fig. \ref{skstar}(a) we show the
\( \Sigma ^{+}K^{*0} \) invariant mass distribution, where the \( K^{*0} \)
is reconstructed in the \( K^{+}\pi ^{-} \) mode. In Fig. \ref{skstar}(b)
we plot the \( \Sigma ^{-}K^{+}\pi ^{+} \) invariant mass distribution where
no signal is evident. A possible explanation for the suppression of the non-resonant
three body decays may be that no quark pairs need to be created in order to
get the \( \Sigma ^{+}K^{*0}(892) \) state (see Fig. \ref{spectator}). 

For the \( \Lambda _{c}^{+}\! \rightarrow \! \Sigma ^{+}K^{+}\pi ^{-} \) sample we
apply a secondary vertex detachment cut, \( l/\sigma _{l}> \)\( 5.5 \), which
rejects much of the combinatoric hadronic background. A minimum cut of \( 50 \)
GeV/$c$  on the momentum of the \( \Lambda _{c}^{+} \) candidate is also
applied. The secondary decay vertex must have a confidence level greater than
\( 10\% \). 

\v{C}erenkov identification cuts are applied to the kaon and pion from the \( \Lambda _{c}^{+} \)
as well as on the charged daughter of the \( \Sigma ^{+} \). In particular
we require \( W_{\pi }-W_{K}>3.5 \) on the \( K^{+} \) while
for the pion (from the \( \Lambda _{c}^{+} \) decay) we require PICON\( >-6 \).
In the \( \Sigma ^{+}\! \rightarrow \! p\pi ^{0} \) case we apply a soft pion-proton
separation cut of \( W_{\pi }-W_{p}>-3 \), while for \( \Sigma ^{+}\! \rightarrow \! n\pi ^{+} \)
the pion must satisfy \( W_{p}-W_{\pi }>-3 \) and PICON\( >-6 \). The pion
from the \( \Sigma ^{+} \) is also required to have a minimum momentum of \( 5 \)
GeV/$c$ . We remove much of the remaining background by requiring \( \sigma _{t}<120 \)\,fs
 for the TS run period and \( \sigma _{t}<150 \)\,fs for the NoTS run period. We also
require the lifetime to be less than six times the \( \Lambda _{c}^{+} \)
lifetime. 

The mass distribution is fit using a double Gaussian plus a linear background. We fixed the
widths, the relative yield ratio and the shift between the means of the two
Gaussians to the Monte Carlo values. The resulting yield is \( 49\pm 10 \) events.

Systematic uncertainties were determined by fit variations on binning, fitting
range, and counting sideband subtracted events in the \( K^{+}\pi ^{-} \)invariant
mass distribution. In order to test for possible biases in the fitting procedure,
we also performed the fit of the \( \Sigma ^{+}K^{+}\pi ^{-} \) mass distribution
using the Monte Carlo shape. We measured the relative branching ratio for statistically
independent sub-samples divided by momentum, particle-antiparticle, run period,
and \( \Sigma ^{+} \) decay modes. Using a technique modeled on the PDG S-factor
method we evaluated a systematic uncertainty from these split samples. After a 
correction for the branching ratio of \( K^{*0}(892)\! \rightarrow \! K^{+}\pi ^{-} \), 
our final result for the branching ratio of \( \Lambda _{c}^{+}\! \rightarrow \! \Sigma ^{+}K^{*0}(892) \)
with respect to \( \Sigma ^{+}\pi ^{+}\pi ^{-} \) is:

$$
\frac{\Gamma (\Lambda _{c}^{+}\! \rightarrow \! \Sigma ^{+}K^{*0}(892))}{\Gamma (\Lambda _{c}^{+}\! \rightarrow \! \Sigma
^{+}\pi ^{+}\pi ^{-})}=(7.8\pm 1.8(stat.)\pm 1.3(syst.))\% 
$$

where the systematic error is obtained by adding in quadrature the contributions from fit 
variants and split samples.

We searched for the similar Cabibbo suppressed decay channel \( \Lambda _{c}^{+}\! \rightarrow \! \Sigma ^{-}K^{+}\pi ^{+} \).
The event selection is identical to that of the \( \Lambda _{c}^{+}\! \rightarrow \! \Sigma ^{+}K^{+}\pi ^{-} \)
selection where the \( \Sigma ^{+} \) is reconstructed in a neutron and a charged
pion. The invariant mass distribution is shown in Fig. \ref{skstar}. To set
an upper limit on the branching ratio for this decay we fit the data using a
double Gaussian with a linear background where all the parameters of the Gaussians,
except for the total yield, are fixed to the Monte Carlo values. The fit returns
a yield of \( 10 \pm 11 \) events. The systematic uncertainty is computed by varying
the range, binning and the fitting shape function. As we do not observe a signal, 
after correcting by the branching ratio of $K^{*0}(892
) \! \rightarrow \! K^+\pi^-$ 
we determine an upper limit of:

$$
\frac{\Gamma (\Lambda _{c}^{+}\! \rightarrow \! \Sigma ^{-}K^{+}\pi ^{+})}{\Gamma (\Lambda _{c}^{+}\! \rightarrow \! \Sigma ^{+}K^{*0}(892))}\leq 35\% 
$$

with a \( 90\% \) confidence level, where we have combined the statistical
and systematic errors in quadrature.

\section{{\normalsize \protect\( \Lambda _{c}^{+}\! \rightarrow \! \Sigma ^{+}K^{+}K^{-}\protect \)decay
mode}\normalsize }

\begin{figure}[htb!]
{\par\centering \resizebox*{8cm}{8cm}{\includegraphics{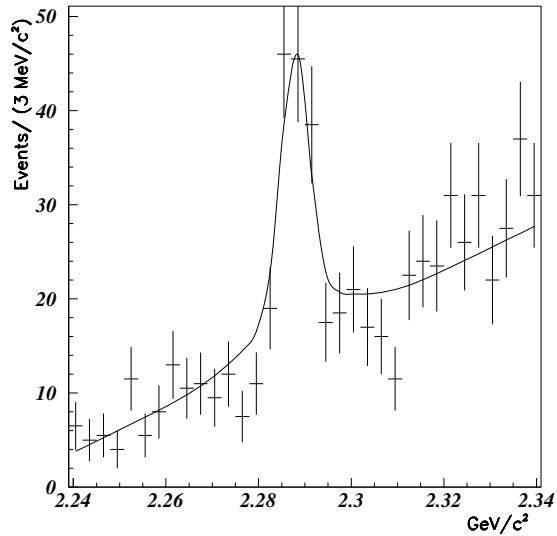}} \par}
\caption{The fit to the \label{skk}\protect\( \Sigma ^{+}K^{+}K^{-}\protect \)
invariant mass distribution fit with a double Gaussian plus a first degree polynomial.}
\end{figure}

The reconstruction of \( \Sigma ^{+}K^{+}K^{-} \)events requires a detachment
cut of \( l/\sigma _{l}> \)\( 3 \), the candidate \( \Lambda _{c}^{+} \)
must have a minimum momentum of \( 30 \) GeV/$c$ and a secondary
vertex with a minimum confidence level of \( 1\% \). The kaons from the \( \Lambda _{c}^{+} \)
must be favored with respect to the pion hypothesis, \( W_{\pi }-W_{K}>1 \).
The identification of the charged pion from the \( \Sigma ^{+} \) (in the $n \pi^+$
decay mode) is achieved by requiring \( W_{p}-W_{\pi }>-3 \) and PICON\( >-6 \).
A soft separation cut of \( W_{\pi }-W_{p}>-3 \), is applied to the proton
for the decay \( \Sigma ^{+}\! \rightarrow \! p\pi ^{0} \). Further, we require the
lifetime resolution \( \sigma _{t}<110 \)\,fs for the TS period and \( \sigma _{t}<140 \)\,fs
 for the NoTS period. The invariant mass distribution is plotted in Fig. \ref{skk}.

We fit the distribution using a double Gaussian for the signal region plus a
linear background and find \( 103\pm 15 \) events. The widths
of the Gaussians, the yield ratio and the shift
between the two means are all fixed to Monte Carlo values. Systematic studies were performed in a manner
similar to that for the \( \Sigma ^{+}K^{*}(892) \) state. Adding in quadrature
the two uncertainties obtained by fit variations and split samples we quote
the branching ratio for \( \Lambda _{c}^{+}\! \rightarrow \! \Sigma ^{+}K^{+}K^{-} \)
relative to \( \Sigma ^{+}\pi ^{+}\pi ^{-} \) to be:

$$
\frac{\Gamma (\Lambda _{c}^{+}\! \rightarrow \! \Sigma ^{+}K^{+}K^{-})}{\Gamma (\Lambda _{c}^{+}\! \rightarrow \! \Sigma^{+}\pi ^{+}\pi ^{-})}=(7.1\pm 1.1(stat.)\pm 1.1(syst.))\%.
$$

\section{{\normalsize \protect\( \Lambda _{c}^{+}\! \rightarrow \! \Sigma ^{+}\phi \protect \)
decay mode}\normalsize }

\begin{figure}[htb!]
{\par\centering \resizebox*{8cm}{8cm}{\includegraphics{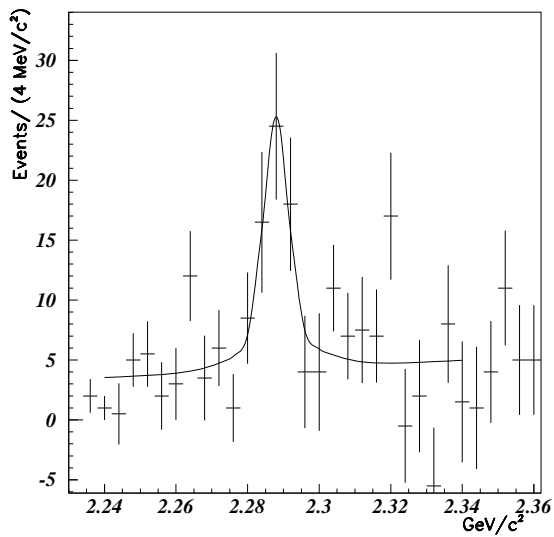}} \par}
\caption{\label{sphi}The figure shows the sideband subtracted invariant mass distribution for \protect\( \Lambda _{c}^{+}\! \rightarrow \! \Sigma ^{+}\phi \protect \)
fit using a double Gaussian for the signal and a linear background. }
\end{figure}

For the \( \Sigma ^{+}K^{+}K^{-} \) final state we also measured the resonant
\( \Sigma ^{+}\phi  \) contribution. The \( \Sigma ^{+}\phi  \) events are
selected using the same cuts used for the \( \Sigma ^{+}K^{+}K^{-} \) inclusive
mode, except: the \( l/\sigma _{l} \) cut is lowered to \( 2.5 \), the \( K^{+}K^{-} \)
invariant mass is required to be within \( 3\sigma  \) (\( 10 \) MeV/$c^{2}$ )
of the \( \phi  \) mass and the absolute difference between the \( \Sigma ^{+}K^{-} \)
invariant mass and the \( \Xi ^{*} \) nominal value (\( 1.690 \) GeV/$c^{2}$ )
must be greater than \( 20 \) MeV/$c^{2}$. This last cut is applied
to suppress the contamination from \( \Lambda _{c}^{+}\! \rightarrow \! \Xi
^{*0}(1690)K^{+} \) where
the \( \Xi ^{*0} \) decays to \( \Sigma ^{+}K^{-} \). A sideband subtraction is 
performed to remove a possible non-resonant contribution. The \( \Sigma ^{+}\phi  \) 
invariant mass distribution, with the \( \Xi ^{*0}(1690) \) exclusion cut, 
is shown in Fig. \ref{sphi}.

The fitting procedure follows the strategy applied in the $\Sigma^+ K^+ K^-$
inclusive mode and give a yield of \( 57\pm 10 \) events. To assess the 
final systematic uncertainty
on this measurement, we follow similar criteria to those described previously. We also 
investigated possible systematic contributions due to our choice of sidebands. 

Adding in quadrature the contribution from fit variations and split samples
and correcting for the branching fraction of \( \phi  \) to \( K^{+}K^{-} \) and
the fraction of events lost from our $\Sigma^+ K^-$ mass cut, 
we quote the final result for the branching ratio of \( \Lambda _{c}^{+}\! \rightarrow \! \Sigma ^{+}\phi  \)
with respect to \( \Lambda _{c}^{+}\! \rightarrow \! \Sigma ^{+}\pi ^{+}\pi ^{-} \)
to be :

$$
\frac{\Gamma (\Lambda _{c}^{+}\! \rightarrow \! \Sigma ^{+}\phi )}{\Gamma (\Lambda _{c}^{+}\! \rightarrow \! \Sigma^{+}\pi ^{+}\pi ^{-})}=(8.7\pm 1.6(stat.)\pm 0.6(syst.))\%
$$

where unseen decay modes of the \( \phi  \) are included.

\section{{\normalsize \protect\( \Lambda _{c}^{+}\! \rightarrow \! \Xi ^{*0}(1690)K^{+}\protect \)
decay mode}\normalsize }

\begin{figure}[!tbh]
{\par\centering \resizebox*{8cm}{8cm}{\includegraphics{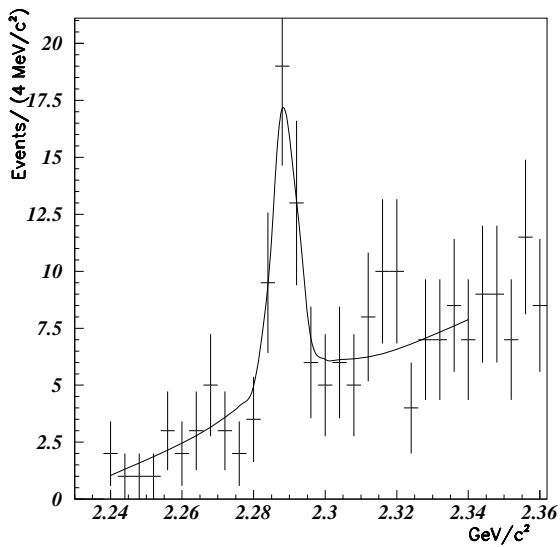}} \par}
\caption{\label{scascdt}Invariant mass distribution for the \protect\( \Lambda _{c}^{+}\protect \)
decay to \protect\( \Xi ^{*0}(1690)(\Sigma ^{+}K^{-})K^{+}\protect \) fit with a double Gaussian for the
signal and a linear background.}
\end{figure}

We also searched for the decay \( \Lambda _{c}^{+}\! \rightarrow \! \Xi ^{*0}(1690)K^{+} \),
with the \( \Xi ^{*0} \) reconstructed in \( \Sigma ^{+}K^{-} \). The cuts used
in the selection of these events are the same as for the inclusive \( \Sigma ^{+}K^{+}K^{-} \)
mode, except that we lowered the detachment cut to \( l/\sigma _{l}> \)\( 2.5 \)
and we applied two mass cuts. The first cut excludes the \( \phi  \) region
by requiring \( M(K^{+}K^{-})>1.03 \) GeV/$c^{2}$. The second cut requires
the \( \Sigma ^{+}K^{-} \) invariant mass to 
be within \( 20 \) MeV/$c^{2}$ of the \( \Xi ^{*0} \) nominal mass of
\( 1.690 \) GeV/$c^{2}$. A clean signal is shown in Fig. \ref{scascdt}. 

We apply the same fitting procedure used in the inclusive
decay mode $\Sigma^+ K^+ K^-$. The signal yield is \( 34\pm 8 \) events. Fit variations 
and split samples were used to evaluate the systematic uncertainty. We quote a final result of:

$$
\frac{\Gamma (\Lambda _{c}^{+}\! \rightarrow \! \Xi ^{*0}(1690)K^{+})}{\Gamma (\Lambda _{c}^{+}\! \rightarrow \! 
\Sigma^{+}\pi ^{+}\pi ^{-})}*B(\Xi ^{*0}(1690)\! \rightarrow \! \Sigma ^{+}K^{-})=(2.2\pm 0.6(stat.)\pm 0.6(syst.))\%. 
$$

From the measurements of the resonant decays in the \( \Sigma ^{+}K^{+}K^{-} \)
final state we infer that almost all of \( \Lambda _{c}^{+}\! \rightarrow \! \Sigma ^{+}K^{+}K^{-} \)
occurs through the resonant modes \( \Sigma ^{+}\phi  \) and \( \Xi ^{*0}(1690)K^{+} \).

\section{{\normalsize Non resonant decay mode \protect\( \Lambda _{c}^{+}\! \rightarrow \! \Sigma ^{+}K^{+}K^{-}\protect \)}\normalsize }

We also looked for a non-resonant contribution to the decay \( \Lambda _{c}^{+}\! \rightarrow \! \Sigma ^{+}K^{+}K^{-} \).
We studied the region \( M(K^{+}K^{-})>1.03 \) GeV/$c^{2}$ and \( M(\Sigma ^{+}K^{-})>1.71 \)
GeV/$c^{2}$ applying the same selection cuts used in the inclusive mode.
\begin{figure}[htb!]
{\par\centering \resizebox*{8cm}{8cm}{\includegraphics{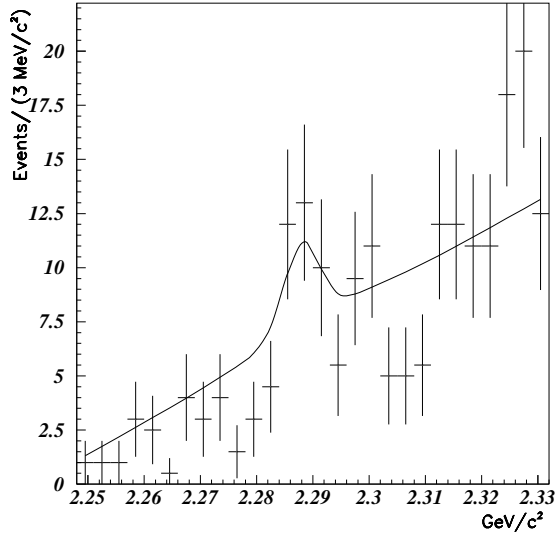}} \par}
\caption{\label{NR}\protect\( \Lambda _{c}^{+}\! \rightarrow \! \Sigma ^{+}K^{+}K^{-}\protect \)
invariant mass distribution for \protect\( M(K^{+}K^{-})>1.03\protect \) GeV/$c^{2}$
and \protect\( M(\Sigma ^{+}K^{-})>1.710\protect \) GeV/$c^{2}$. The fit is performed
using a double Gaussian for the signal plus a linear background.}
\end{figure}

In Fig. \ref{NR} we show the \( \Sigma ^{+}K^{+}K^{-} \) invariant mass
distribution where the double Gaussian fitting procedure has been applied. All
the parameters of the double Gaussian, except the total yield, are fixed to their
Monte Carlo values. The yield from the fit is \( 14\pm 8 \) events. After 
further corrections due to
possible contamination from \( \Lambda _{c}^{+}\! \rightarrow \! \Sigma ^{+}\phi  \)
and \( \Lambda _{c}^{+}\! \rightarrow \! \Xi ^{*0}(1690)K^{+} \) decays we find a
yield of \( 8\pm 8 \) events. With no evidence of a signal, we quote the \( 90\% \)
confidence level limit for the non-resonant component of the decay \( \Lambda _{c}^{+}\! \rightarrow \! \Sigma ^{+}K^{+}K^{-} \)
with respect to \( \Lambda _{c}^{+}\! \rightarrow \! \Sigma ^{+}\pi ^{+}\pi ^{-} \)
to be:

$$
\frac{\Gamma (\Lambda _{c}^{+}\! \rightarrow \! \Sigma ^{+}K^{+}K^{-})_{NR}}{\Gamma (\Lambda _{c}^{+}\! \rightarrow \! \Sigma ^{+}\pi ^{+}\pi ^{-})}<2.5\%.
$$

Our systematic error, added in quadrature to the statistical error
and included in the upper limit, is determined by varying the fit conditions
in a manner similar to that previously described.

\section{{\normalsize Conclusions}\normalsize }

\begin{table}[htb!]
\begin{center}

\caption{FOCUS results compared to previous measurements \cite{Abe:2001mb,Avery:1993vj} where applicable. The
relative efficiencies are computed with respect to the normalization mode.}
\label{tb:results}
\begin{tabular}{|c | c | c | c|c | }

\hline\hline


               & Efficiency               & FOCUS results      & BELLE results & CLEO results\\ 
               & Ratio               &      & & \\ \hline

$\frac{\Gamma (\Lambda_{c}^{+} \! \rightarrow \! \Sigma^{+} K^{*}(892))}{\Gamma (\Lambda_{c}^{+} \! \rightarrow \!
 \Sigma^{+} \pi^{+} \pi ^{-}) }$ & 0.35 & $(7.8 \pm 1.8 \pm 1.3)\% $	   &- &-\\ \hline

$\frac{\Gamma (\Lambda _{c}^{+}\! \rightarrow \! \Sigma ^{-}K^{+}\pi ^{+})}{\Gamma (\Lambda_{c}^{+}
\! \rightarrow \! \Sigma ^{+}K^{*}(892))}$	  & 1.49 & $<35\%$ @ $90\%$ $CL$ & - &-\\ \hline

$\frac{\Gamma (\Lambda _{c}^{+} \! \rightarrow \! \Sigma ^{+}K^{+}K^{-})}{\Gamma(\Lambda _{c}^{+} \! \rightarrow \!
\Sigma ^{+}\pi ^{+}\pi ^{-}) }$ & 0.85 & $(7.1 \pm 1.1 \pm 1.1)\%$ & $(7.6 \pm 0.7 \pm 0.9)\%$ & $(9.5 \pm 1.7 \pm 1.9)\%$ \\\hline

$\frac{\Gamma (\Lambda _{c}^{+}\! \rightarrow \! \Sigma ^{+} \phi)}{\Gamma (\Lambda _{c}^{+}\! \rightarrow \!
\Sigma ^{+}\pi ^{+}\pi ^{-}) }$ & 0.39 & $(8.7\pm 1.6\pm 0.6)\%$ & $(8.5 \pm 1.2 \pm 1.2)\%$ &$(9.3 \pm 3.2 \pm 2.4)\%$  \\\hline

$\frac{\Gamma (\Lambda _{c}^{+}\! \rightarrow \! \Xi ^{*}(\Sigma^+K^-) K^+)}{\Gamma (\Lambda _{c}^{+}\! \rightarrow \!
\Sigma ^{+}\pi ^{+}\pi ^{-}) }$ & 0.92 & $(2.2 \pm 0.6 \pm 0.6)\% $ & $(2.3 \pm 0.5 \pm 0.5)\%$ &-\\ \hline

$\frac{\Gamma (\Lambda _{c}^{+}\! \rightarrow \! \Sigma^+ K^- K^+)_{NR}}{\Gamma (\Lambda _{c}^{+}\! \rightarrow \!
\Sigma ^{+}\pi ^{+}\pi ^{-}) }$ & 0.44 & $<2.8\%$ @ $90\%$ $CL$  & $<1.8\%$ @ $90\%$ $CL$  &-\\ \hline

\hline
\hline
\end{tabular}
\end{center}
\end{table}

We have measured the branching ratio of four \( \Lambda _{c}^{+} \)
decay modes containing a \( \Sigma ^{+} \) particle reconstructed in both the
\( p\pi ^{0} \) and \( n\pi ^{+} \) channels. These modes are
\( \Lambda _{c}^{+}\! \rightarrow \! \Sigma ^{+}K^{*0}(892) \) , \( \Lambda _{c}^{+}\! \rightarrow \! \Sigma ^{+}K^{+}K^{-} \)
inclusive, \( \Lambda _{c}^{+}\! \rightarrow \! \Sigma ^{+}\phi  \) and \( \Lambda _{c}^{+}\! \rightarrow \! \Xi ^{*0}(\Sigma ^{+}K^{-})K^{+} \).
For these last two modes our measurements are consistent with the recent results
reported by the Belle collaboration \cite{Abe:2001mb}. We also set an upper limit for the Cabibbo
suppressed decay mode \( \Lambda _{c}^{+}\! \rightarrow \! \Sigma ^{-}K^{+}\pi ^{+} \)and
the non-resonant contribution to \( \Lambda _{c}^{+}\! \rightarrow \! \Sigma ^{+}K^{+}K^{-} \). Our
final results, the relative efficiency of each mode with respect to the normalization mode as well as
comparisons to previous measurements, are summarized in Table \ref{tb:results}.

\section{Acknowledgements}
We wish to acknowledge the assistance of the staffs of Fermi National
Accelerator Laboratory, the INFN of Italy, and the physics departments of the
collaborating institutions. This research was supported in part by the U.~S.
National Science Foundation, the U.~S. Department of Energy, the Italian
Istituto Nazionale di Fisica Nucleare and Ministero dell'Universit\`a e della
Ricerca Scientifica e Tecnologica, the Brazilian Conselho Nacional de
Desenvolvimento Cient\'{\i}fico e Tecnol\'ogico, CONACyT-M\'exico, the Korean
Ministry of Education, and the Korean Science and Engineering Foundation.

\bibliographystyle{myapsrev}
\bibliography{lc_paper_plb}

\end{document}